\documentclass[article,aps, amsfonts, nofootinbib]{revtex4}

\usepackage{epsf}
\usepackage{amscd}
\usepackage{amsmath}
\input xy
\xyoption{all} 
\usepackage[usenames,dvips]{color}  

\newcommand{\beq}{\begin{equation}}
\newcommand{\eeq}{\end{equation}}
\newcommand{\bea}{\begin{eqnarray}}
\newcommand{\eea}{\end{eqnarray}}

\newcommand{\nn}{\nonumber}
\newcommand{\benn}{\begin{displaymath}}
\newcommand{\eenn}{\end{displaymath}}

\def\slashchar#1{\ensuremath{                               %
   \setbox0=\hbox{${}#1{}$}       
   \dimen0=\wd0                                 
   \setbox1=\hbox{/} \dimen1=\wd1               
   \ifdim\dimen0>\dimen1                        
      \rlap{\hbox to \dimen0{\hfil/\hfil}}      
      {}#1{}                                    
   \else                                        
      \rlap{\hbox to \dimen1{\hfil${}#1{}$\hfil}}   
      /                                         
   \fi}}                                        %

\begin{document}

\title{Restless pions from orbifold boundary conditions:\\
 an explicit construction for noise reduction in lattice QCD } 
\author{Paulo F.~Bedaque\footnote{{\tt bedaque@umd.edu}}}
\affiliation{University of Maryland, College Park, MD, USA} 
\author{Andre Walker-Loud} 
\email[]{walkloud@wm.edu}
\affiliation{University of Maryland, College Park, MD, USA} 
\affiliation{College of William and Mary, Williamsburg, VA, USA}

\preprint{}
\begin{abstract}
  The exponentially decreasing signal to noise ratio in multibaryon correlators is the main obstacle to a first principles, QCD-based calculation of the nuclear force. Recently, we have proposed an orbifold boundary condition (``restless pions") that can dramatically improve this matter. Here we develop the idea further by proposing an explicit algorithm that can be used with purely periodic, ``off the shelf" gauge configurations. We also discuss finite volume corrections with the new boundary conditions and the use of the ``L\"{u}scher formula'' for the phase shifts.
\end{abstract}
\maketitle

\section{Introduction}

A first principles, QCD-based understanding of the origins of the nuclear force is still lacking.  Besides the obvious theoretical interest, such a calculation would open the door for  first principles calculations of other multi-baryon quantities whose values are unknown and/or difficult to be measured and that impact nuclear, particle and astrophysics. A step forward in this direction was taken by recent unquenched calculations of baryon-baryon low energy phase shifts \cite{Beane:2006mx,Beane:2006gf}. They revealed what is currently the main difficulty facing this program: the exponential decrease with (euclidean) time of the signal to noise ratio in multi-baryon correlators. We suggested recently \cite{Bedaque:2007pe} a method to alleviate this problem.  It can be viewed as the use of a special kind of boundary condition  along the spatial directions generated by an orbifold condition on the quark fields. One drawback of the method is that it leads to a non-positive fermion determinant. Thus, the only way  it can be used in practice is by combining the new boundary conditions on the valence quarks while keeping standard periodic boundary conditions for the sea quarks. While this raises some conceptual issues, from the practical point of view the use of  mixed boundary conditions is convenient as the most costly part of a numerical lattice calculation -- the generation of gauge configurations -- does not have to be redone with the new boundary conditions.
 The purpose of the present paper is to make explicit the algorithm using previously generated periodic lattices and to discuss several issues raised, such as finite volume corrections and the validity of the L\"{u}scher formula~\cite{hamber_et_al,luscher_1,luscher_2}.
 
 The  origin of the noise  increase with time  in lattice calculations is well understood \cite{lepage,Bedaque:2007pe}. A simple analysis shows that the signal-to-noise ratio of a two-nucleon correlator is given by
 
\beq\label{eq:B_signaltonoise}
	\frac{C(t)}{\sqrt{\frac{1}{N}\sigma_C^2(t)}}\ \stackrel{\small t\rightarrow\infty}{\longrightarrow}\ 
		A\sqrt{N} \frac{e^{-3E_M t}}{e^{-3E_\pi t}} 
		\sim \sqrt{N} e^{-(2E_M-3 E_\pi) t}\, ,
\eeq where $C(t)=\langle q^3(t)q^3(t) \bar q^3(0)\bar q^3(0) \rangle$ is the two-nucleon correlator (spin, isospin and spatial indices are omitted for simplicity),  $\sigma_C^2(t)$ the variance of  $C(t)$  after $N$ measurements. $E_M$ and $E_\pi$ are the lowest energies allowed for the nucleon and pion in the lattice. With periodic boundary conditions zero momentum states for both the nucleon and the pion exist and the minimum energy equals their rest mass, $E_M= M$, $E_\pi=m_\pi$. The basic idea employed in \cite{Bedaque:2007pe} is to find a boundary condition where the pion zero-modes are forbidden resulting in $E_M=M$ but $E_\pi > m_\pi$. It is not obvious how to accomplish this since only the boundary conditions of the quark and gluon fields, not the hadron fields, are chosen at will. The solution found in \cite{Bedaque:2007pe} is to formally extend the lattice in one direction from $0 < z < L$ to $-L < z < L$ and impose the parity orbifold condition

\bea\label{eq:1d_qg}
q(x,y,-z,t) &=& \mathcal{P}_z q(x,y,z,t),\nn\\
A_3(x,y,-z,t) &=& - A_3(x,y,z,t),\nn\\
A_\mu(x,y,-z,t) &=& A_\mu(x,y,z,t), \ {\mu\neq 3},
\eea where $ \mathcal{P}_z=i \gamma^5\gamma^3$ implements a z-axis reflection. In other words, the fields in the $z<0$ region are not independent variables but merely the parity reflection of the fields in the $z>0$ region. The orbifold conditions for the quark and gluon fields imply in orbifold condition for the pion  fields

\beq\label{eq:pion_orbifold}
\pi(x,y,-z,t) = -\pi(x,y,z,t).
\eeq   Equation \eqref{eq:pion_orbifold} forbids the pion zero mode $\pi(x,y,z,t)=$ constant.

Since the fields in the $z<0$  are related to the ones in the $z>0$ region by a symmetry of QCD (parity), the action in both regions is the same, for every field configuration. Thus, there is no need to actually double the size of the lattice in an actual simulation. The only effect of the orbifold condition in equation~(\ref{eq:1d_qg}) is right at the boundary at $z=0$. The precise form of the action at the boundary depends on the particular regularization being used. For instance, if we index the sites in the z-direction by $z=1,\cdots, L$ and $z=-L,\cdots , -1$ (no $z=0$ site!) and Wilson quarks are used we have

\bea\label{eq:lattice_orbifold}
S &=&\kappa \left[\bar q_{-1} (\gamma_3-r) q_1 - \bar q_{1} (\gamma_3+r) q_{-1} \right] +(\bar q_1 q_1+\bar q_{-1} q_{-1})+\cdots\nn\\
&=& -2 \kappa \bar q_{1} (\gamma_3+r) \mathcal{P}_z q_1 +2 \bar q_1 q_1+\cdots,
\eea 
where $\kappa$ is the hopping parameter, the index on the quark fields denotes the position in $z$ (the remaining coordinates are implicit) and the dots denote the contributions from the two sides of the bulk, $z>0$ and $z<0$ (which are equal to each other). We see then that the orbifolded $[-L,L]$ lattice is equivalent to a $[1,L]$ lattice with some extra terms residing at the boundary, as is the case with any  boundary condition.%
\footnote{Notice that, contrary to the continuum case, the boundary conditions in lattice field theory are already contained in the action. Different lattice action terms localized at the boundary imply different boundary conditions in the continuum and the relation between them is, in general, a complicated dynamical question.} Despite the fact that the $z<0$ region is not actually present in the simulations, we will use the extended lattice as a theoretical device in order to better analyze the properties of our construction.
 
 The orbifold condition of equation~(\ref{eq:1d_qg}) (that we call ``one-dimensional orbifold'') is inconvenient because the spin up nucleon field zero modes are also eliminated. A slight change (that we call ``three-dimensional orbifold'') leads to a more manageable method:
 \bea\label{eq:3d_qg}
q(-x,-y,-z,t) &=& \mathcal{P} q(x,y,z,t),\nn\\
A_4(-x,-y,-z,t) &=&  A_4(x,y,z,t),\nn\\
A_\mu(-x,-y,-z,t) &=& A_\mu(x,y,z,t), \ {\mu\neq 0},
\eea where $ \mathcal{P}=\gamma^4$ implements a parity reflection. The pion and nucleon fields satisfy

\bea
\pi(-x,-y,-z,t) &=& -\pi(x,y,z,t),\nn\\
N(-x,-y,-z,t) &=& N(x,y,z,t),
\eea where in the last line we assumed the nucleons to be non-relativistic and used the non-relativistic limit $\gamma^4\to 1$.

As mentioned above, the orbifold boundary conditions cannot be used for the sea quarks due to the non-positivity of the determinant. Our proposal avoids this issue by using  mixed boundary conditions: orbifold boundary conditions for the valence quarks and periodic for the sea quarks (and gluons). That means that periodic gauge configurations can (and have to) be used. That raises the question of how the orbifold condition can meld with periodicity. We will analyze this question in section II for both the one-dimensional and the three-dimensional orbifold and determine the allowed pion modes. In section III we will discuss the finite volume corrections in the orbifold construction and the use of the L\''{u}scher formula.

\section{Explicit lattice construction }

\subsection{``One-dimensional" parity orbifold}
For clarity, in this section we considered the simpler but less useful construction where the new boundary condition is imposed in only on direction.

\begin{figure}[!htbp]
  \centerline{{\epsfxsize=2.0in \epsfbox{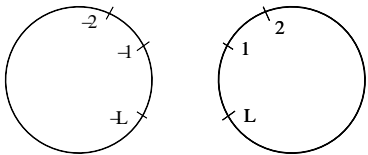}}}
\noindent
\caption{Lattice where the gauge and sea quarks ``live". Only the z-coordinate is shown. The fields on the left sub-lattice are not independent but only a parity reflection of the ones on the right sub-lattice.}
\label{fig:sea_orb}
\end{figure}

Consider a gauge configuration generated using standard periodic boundary condition for both gluons and (sea) quarks and index the sites on the z-direction by $z=1,2,\cdots,L$. We can formally extend this configuration to one defined for $z=1,2,\cdots,L$ and $z=-L,\cdots,-1$ by defining the gauge links $U_\mu$ in the $z<0$ region by

\bea
U_z(x,y,z,t) &=& U_z^\dagger(x,y,-z,t),\nn\\
U_\mu(x,y,z,t) &=& U_\mu(x,y,-z,t), \ \ \text{for}\ \ \mu\neq 3.
\eea An ensemble of gauge configurations generated by this procedure is identical to one that would be generated by actually simulating a larger $-L\le z\le L$ lattice where the fields in the two halves are constrained to obey the orbifold condition:

\bea\label{eq:1d_seaorbifold}
q_s(x,y,-z,t) &=& \mathcal{P}_z q_s(x,y,z,t),\nn\\
A_3(x,y,-z,t) &=& - A_3(x,y,z,t),\nn\\
A_\mu(x,y,-z,t) &=& A_\mu(x,y,z,t), \ {\mu\neq 3},
\eea where $q_s$ are the (sea) quark fields. Notice that there is no gauge link connecting the $z=1$ and $z=-1$ sites (see figure \ref{fig:sea_orb}.

\bigskip

\begin{figure}[!htbp]
  \centerline{{\epsfxsize=2.0in \epsfbox{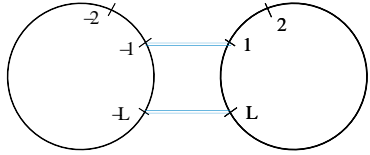}}}
\noindent
\caption{Lattice where the gauge and valence quarks ``live". Only the z-coordinate is shown. The fields on the left sub-lattice are not independent but only a parity reflection of the ones on the right sub-lattice. The double line denotes a hopping term with trivial gauge field $U_z=\openone$.}
\label{fig:valence_orb}
\end{figure}

The action for the valence quarks is also defined on a doubled lattice. The difference from the gauge and sea quark sector is that
  hopping terms are added connecting the $z=1$ to $z=-1$  and $z=L$ and $z=-L$ for all values of $x$, $y$ and $t$. In these links the gauge field is defined to be trivial ($U_z=\openone$). This construction is pictured on figure \ref{fig:valence_orb}. The valence quark fields at $z<0$ are, again, not independent of their counterparts at $z>0$; $q_v(x,y,-z,t)$ is constrained to be the parity reversal of $q_v(x,y,z,t)$
  
\beq\label{eq:eq:1d_valenceqg}
q_v(x,y,z,t) = \mathcal{P}_z q_v(x,y,z,t).
\eeq Notice that, in the valence action, the nodes with $z=1$, $z=L$ have one more nearest neighbor that the remaining sites.

\bigskip

The new plaquettes in the  action involving the links between $z=1$ and $z=-1$ ($z=L$ and $z=-L$ ) are trivial and don't contribute to the action. The action in the $z<0$ sub-lattice, for any field configuration, is the same as the one in the $z>0$ sub-lattice, on account of the invariance of the lattice QCD action under parity transformations. Thus, in an actual simulation, only the $z>0$ sub-lattice is used. The only effect of our construction in the valence action is the addition of an extra term at the boundary at $z=1$. The precise form of this term is determined by the particular lattice action used and the orbifold condition in eq.~(\ref{eq:1d_qg}). For the Wilson action, for instance, we have the result in equation~(\ref{eq:lattice_orbifold}).

In the continuum limit, the orbifold conditions in eq.~(\ref{eq:1d_seaorbifold},\ref{eq:eq:1d_valenceqg}) imply in

\bea\label{eq:bc_orbifold1}
q_s(x,y,-z,t) &=& \mathcal{P}_z q_s(x,y,z,t),\nn\\
q_v(x,y,-z,t) &=& \mathcal{P}_z q_v(x,y,z,t),\nn\\
A_\mu(x,y,-z,t) &=& A_\mu(x,y,z,t),\ {\rm for}\ \mu\neq 3\nn\\
A_3(x,y,-z,t) &=& -A_3(x,y,z,t),\ {\rm for}\ \mu=3.
\eea The sea quark fields are periodic along the z-direction (going from $z=+0$ to $z=L$ or from $z=-0$ to $z=-L$) but can be discontinuous in going from $z=+0$ to $z=-0$ since there is no hopping term connecting the two halves of the extended lattice. The valence quarks, however, are also continuous along the $z=+0$ to $z=-0$ border.

We can now use equation (\ref{eq:bc_orbifold1})  to figure out the boundary conditions satisfied by the hadrons in the continuum limit. In the spirit of partially quenched chiral perturbation theory \cite{Sachrajda:2004mi,Bedaque:2004ax,Tiburzi:2005hg}, there are three kinds of pions, depending whether their constituents quarks and antiquarks are sea or valenve quarks. For all three kinds of pions we have the parity orbifold condition:
\bea
\pi_{vv}(x,y,-z,t) &=& -\pi_{vv}(x,y,z,t),\nn\\
\pi_{vs}(x,y,-z,t) &=& -\pi_{vs}(x,y,z,t),\nn\\
\pi_{ss}(x,y,-z,t) &=& -\pi_{ss}(x,y,z,t).
\eea 
Furthermore, the translation invariance in the sea sector implies the periodicity of the $\pi_{ss}$ field in all three spatial directions. That is, not only $\pi_{ss}(+0,y,z,t) = \pi_{ss}(L,y,z,t)$, $\pi_{ss}(-0,y,z,t) = \pi_{ss}(-L,y,z,t)$ (and similarly in the $y$ and $z$ directions) but its derivatives are also the same ( $\partial_x\pi_{ss}(+0,y,z,t) = \partial_x\pi_{ss}(L,y,z,t)$, $\cdots$ ). However, $\pi_{ss}(+0,y,z,t)$ is not necessarily the same as $\pi_{ss}(-0,y,z,t)$. Thus, the condition $\pi_{ss}(x,y,-z,t) = -\pi_{ss}(x,y,z,t)$ does {\it not} imply in  $\pi_{ss}(x,y,0,t)=0$. 

On the other hand, the continuity of $q_v$ at $z=0$ (the result of the extra links) implies in
\beq
\pi_{vv}(x,y,L,t) = \pi_{vv}(x,y,+0,t)=-\pi_{vv}(x,y,-0,t) =\pi_{vv}(x,y,-L,t) = 0.\\
\eeq The vanishing of $\pi_{vv}$ at $z=0$ eliminates its zero mode which is the main point of our construction. $\pi_{sv}$ satisfies the same boundary conditions as $\pi_{ss}$.

 The boundary conditions discussed above specify the allowed modes for each kind of pion:
\bea
\pi_{vv}(z) &\sim&  e^{ i\frac{2\pi}{L}(n_x x +n_y y)}\ \sin\left(\frac{n_z\pi z}{L}\right),\ \ n_x, n_y = 0, \pm 1, \pm 2, \cdots, \ n_z=1,2, \cdots\\ 
\pi_{ss}(z), \pi_{vs}(z) &\sim& 
e^{ i\frac{2\pi}{L}(n_x x +n_y y)}
\begin{cases}
{\rm sign}(z)\cos\left(\frac{2n_z\pi z}{L}\right)\\
\sin\left(\frac{2n_z\pi z}{L}\right)
\end{cases}
,\ \ 
n_x, n_y = 0, \pm 1, \pm 2, \cdots, \,\ \ n_z=0,1,2, \cdots .
\eea 
The discontinuity for $\pi_{ss}, \pi_{sv}$ may seem strange. Remember, however, that there is no link between $z>0$ and $z<0$ in the sea quark action, so the discontinuity costs no energy. 
It is important to note that the presence of zero modes for $\pi_{vs}$ and $\pi_{ss}$ does not spoil the noise reduction of baryon correlators.  This is because the sea-sea and sea-valence pions can only appear in intermediate states {\it addition} to the valence pions.%
\footnote{Notice that in correlators of two (valence) nucleons  will always include, at least, 6 valence pions and thus the minimal energy of allowed intermediate states will be bounded from below by (six times) the minimal energy of the {\it valence} pions. }

The valence nucleons (those composed of three valence quarks) are periodic in the $x$ and $y$ directions, continuous at $z=0$ and satisfy the orbifold condition $N(x,y,-z,t) = \mathcal{P}_z N(x,y,z,t)$. At $z=0$ and in the non-relativistic limit this implies in
\beq
N(x,y,0,t) = \sigma_z N(x,y,0,t).
\eeq The up and down spin components obey different boundary conditions.

\subsection{``Three-dimensional" parity orbifold}
 This case is very similar to the ``one-dimensional" orbifold case. Again, we start from  gauge configurations periodic in all three directions. This time, however, we choose  the $x$ and $y$ coordinates to vary from $-L/2$ to $L/2$ while the $z$ coordinate varies from $1$ to $L$.   We  formally extend the gauge configurations to the $z<0$ region through the relations:
 
 \bea
q_s(-\mathbf{r},t) &=& \mathcal{P} q_s(\mathbf{r},t),\nn\\
\mathbf{A}(-\mathbf{r},t) &=&- \mathbf{A}(\mathbf{r},t),\nn\\
A_4(-\mathbf{r},t) &=& A_4(\mathbf{r},t),
\eea with $\mathcal{P}=\gamma_4$ is the usual parity operator. The valence quark propagators are computed in this background after trivial ($U=\openone$) links are added connecting $(x,y,1,t)$ to $(-x,-y,-1,t)$. Again, the result of having the extended lattice is simply an extra term in the lattice action living at the first node $z=1$ whose precise form depends on the fermion discretization used. Contrary to the ``one-dimensional'' orbifold case though, this extra term is non-local and corresponds to hops from $(x,y,1,t)$ to/from $(-x,-y,1,t)$. This construction is equivalent as having valence quark fields that are periodic in the $x$, $y$ and $z$ directions and obey the orbifold condition
\beq\label{eq:3d_orbifold}
q_v(-\mathbf{r},t) = \mathcal{P} q_v(\mathbf{r},t).\\
\eeq 
In addition (and contrary to the sea quarks), $q_v$ is continuous in going from $(x,y,+0,t)$ to $(-x,-y,-0,t)$.  Using the Wilson lattice action as an example, an explicit construction of the valence quark action, we have%
\footnote{Recall that there is no $z=0$ site, for simplicity of writing down the action.  Also, there is a similar contribution to the action at $z=L$.} 
\begin{align}
S = \kappa \left[
	\bar{q}_{x,y,1,t} (\gamma_3 +r) q_{x,y,-1,t}
	-\bar{q}_{x,y,-1,t} (\gamma_3 - r) q_{x,y,1,t}
	\right]
	+\left( \bar{q}_{x,y,1,t}\, q_{x,y,1,t} + \bar{q}_{x,y,-1,t}\, q_{x,y,-1,t} \right)
	+\cdots
\end{align}
where as in eq.~\eqref{eq:lattice_orbifold}, the ellipses denote the action from the two sides of the bulk.  Using the three-dimensional orbifolding conditions, eq.~\eqref{eq:3d_orbifold}, one can show that the action is given by
\begin{align}\label{eq:3Dorbifold}
S &= \kappa \left[
	\bar{q}_{x,y,1,t} (\gamma_3 +r) \mathcal{P} q_{-x,-y,1,t}
	-\bar{q}_{-x,-y,1,t} \mathcal{P} (\gamma_3 - r) q_{x,y,1,t}
	\right]
	- \left[ \bar{q}_{x,y,1,t}\, q_{x,y,1,t} + \bar{q}_{-x,-y,1,t}\, \mathcal{P}^2\, q_{-x,-y,1,t} \right]
	+\cdots
\nonumber\\
	&= 2\kappa\,  
		\bar{q}_{x,y,1,t} (\gamma_3 +r) \mathcal{P} q_{-x,-y,1,t}
		-2 \bar{q}_{x,y,1,t}\, q_{x,y,1,t}
	+\cdots
\end{align}
where the second line comes from a relabeling of the dummy spatial indices under summation.  Notice the differences between this boundary action and that for the one-dimensional orbifolding construction, eq.~\eqref{eq:lattice_orbifold}.  In the one-dimensional case, the extra boundary action is entirely local while in the three-dimensional case, eq.~\eqref{eq:3Dorbifold}, the boundary action is explicitly non-local.

We can now determine the boundary conditions satisfied by the hadronic fields in the continuum limit. Like in the ``one-dimensional'' orbifold

\bea
\pi_{vv}(-\mathbf{r},t) &=& -\pi_{vv}(\mathbf{r},t),\nn\\
\pi_{vs}(-\mathbf{r},t) &=& -\pi_{vs}(\mathbf{r},t),\nn\\
\pi_{ss}(-\mathbf{r},t) &=& -\pi_{ss}(\mathbf{r},t).\nn\\
\eea

The valence pions, besides obeying the same boundary/orbifold conditions as the sea and sea-valence pions,  are also continuous along the $z=0$ plane. This implies on the vanishing of $\pi_{vv}$ at the origin $(0,0,0,t)$ since 
$\pi_{vv}(0,0,0,t) = -\pi_{vv}(0,0,0,t) = 0$.  Due to the periodicity, $\pi_{vv}$ then vanishes at the border of the lattice as can be seen through the following relations,
\bea
\pi_{vv}(L/2,0,0,t) &=& -\pi_{vv}(-L/2,0,0,t) = -\pi_{vv}(L/2,0,0,t) =0,
\eea 
and similarly for the $y$ and $z$ directions.  The first relation arises from the parity-orbifold construction while the second relation arises from the periodicity of the lattice.

\begin{figure}[!htbp]
  \centerline{{\epsfxsize=3.5in \epsfbox{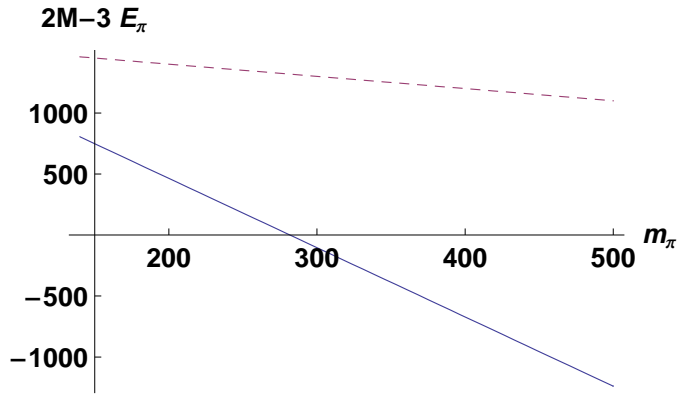}}}
\noindent
\caption{Estimated energy splitting which controls the decay of the signal-to-noise for the two nucleon system, $2M - 3E_\pi$.  The (red) dashed line is for periodic boundary conditions with $E_\pi = m_\pi$ while the blue line is for the three dimensional orbifold, with $E_\pi = \sqrt{9(\pi/L)^2 + m_\pi^2}$, in which we have held $m_\pi L=4$ fixed in our estimation.  The pion mass dependence of the nucleon mass was estimated from Refs.~\cite{WalkerLoud:2008bp,WalkerLoud:2008pj}.}
\label{fig:noise_estimate}
\end{figure}

Taking into account the boundary conditions satisfied by the $\pi_{vv}$, $\pi_{sv}$ and $\pi_{ss}$ fields we find that the allowed modes are
\bea
\pi_{vv}(x,y,z) &\sim& \sin(\frac{2n_x\pi x}{L}) \sin(\frac{2n_y\pi y}{L}) \sin(\frac{n_z\pi z}{L}), \nn\\
\pi_{ss}(x,y,z), \pi_{sv}(x,y,z) &\sim& 
  \sin(\frac{2n_x\pi x}{L}) \sin(\frac{2n_y\pi y}{L}) \begin{cases}
                                                  {\rm sign}(z)\cos(\frac{n_z\pi z}{L})\\ \\ 
                                                   \sin(\frac{n_z\pi z}{L})
                                                   \end{cases},\nn \\
&&\cos(\frac{2n_x\pi x}{L}) \cos(\frac{2n_y\pi y}{L}) \begin{cases}
                                                  {\rm sign}(z)\cos(\frac{n_z\pi z}{L})\\ \\ 
                                                   \sin(\frac{n_z\pi z}{L})
                                                   \end{cases},\nn \\
 &&\sin(\frac{2n_x\pi x}{L}) \cos(\frac{2n_y\pi y}{L})  \cos(\frac{n_z\pi z}{L}),\nn \\
 &&\cos(\frac{2n_x\pi x}{L}) \sin(\frac{2n_y\pi y}{L})  \cos(\frac{n_z\pi z}{L}) .                                        \eea Notice that the minimal energy of a valence pion is $\sqrt{2(2\pi/L)^2 +(\pi/L)^2 +m_\pi^2}$, a value larger than the one quoted in \cite{Bedaque:2007pe} where the appropriate boundary conditions at $z=L$ were not taken into account; see figure~\ref{fig:noise_estimate} for an updated estimate of the improvement in the signal to noise.

Nucleon fields satisfy, as a consequence of equation (\ref{eq:3d_orbifold}), the boundary condition
$N(-\mathbf{r},t) = \gamma^4 N(\mathbf{r},t) $. In the non-relativistic limit this reduces to $N(-\mathbf{r},t) = N(\mathbf{r},t)$. Thus, unlike pions, the nucleon zero mode is allowed. This feature is very convenient to construct two-nucleon states at rest and use the L\"{u}scher formula to extract the zero energy scattering phase shift.

\bigskip

\section{Finite volume corrections}
The finite volume corrections to quantities computed in the lattice are, as a long distance property, dependent on the boundary conditions used. In periodic lattices, masses, decay constants and scattering lengths receive only exponentially suppressed finite volume corrections proportional to $\sim e^{-m_\pi L}$. Unfortunately, the absence of pion zero modes enhances these finite volume corrections and they become only power law suppressed. It is simple to estimate these effects. For observables describable by a low energy effective theory like chiral perturbation theory these effects are dominated by pion loops. The difference between a one-loop diagram computed at finite volume and in the infinite volume limit is of the form
\beq
\Delta \mathcal{A} = \frac{1}{L^3} \sum_q f(\vec{q}) - \int \frac{d^3q}{(2\pi)^3} f(\vec{q}),
\eeq for some function $f(q)$,  where the sum is over the allowed momenta for the pions. In periodic lattices $q=2\pi \vec{n}/L$ with $\vec{n}$ having integers components. In this case it can be shown that, as long as $f(q)$ has no singularity along the real axis (that is, the particles in the loop cannot be on-shell simultaneously), $\Delta\mathcal{A}\sim e^{-m_\pi L} $. When the allowed momenta do not include the zero mode $\Delta \mathcal{A}$ can be larger. Let us take, for instance, the case of the ``three-dimensional'' orbifold. The sum over the allowed valence pion modes is

\bea
\Delta \mathcal{A} &=& \frac{1}{L^3} \sum_{n_x,n_y,n_z=1,2,\cdots} 
f\left(\frac{2\pi n_x}{L},\frac{2\pi n_y}{L},\frac{\pi n_z}{L}\right)
 - \int \frac{d^3q}{(2\pi)^3} f(\vec{q})\nn\\
 &=&\frac{1}{2L^3} \sum_{n_x,n_y,n_z=\pm 1,\pm 2,\cdots} 
f\left(\frac{2\pi n_x}{L},\frac{2\pi n_y}{L},\frac{\pi n_z}{L}\right)
 - \int \frac{d^3q}{(2\pi)^3} f(\vec{q})\nn\\
 &=&\frac{1}{2L^3} \sum_{n_x,n_y,n_z=0,\pm 1,\pm 2,\cdots} 
f\left(\frac{2\pi n_x}{L},\frac{2\pi n_y}{L},\frac{\pi n_z}{L}\right)
 - \int \frac{d^3q}{(2\pi)^3} f(\vec{q})
 -\frac{1}{2L^3} f(0,0,0)\nn\\
&=&\mathcal{O}(e^{-m_\pi L}) -\frac{1}{2L^3} f(0,0,0),
 \eea where we assumed that $f(\vec{q})$ is an even function. The $f(0,0,0)/L^3$ term is the power law finite volume correction mentioned above. 
 
 Through a more detailed analysis one can estimate the numerical value of these power law corrections. Since this analysis is dependent on the particular observable of interest we will refrain from pursuing it here.  However, we note that because of chiral symmetry, the pions, at leading order, are derivatively coupled to hadrons and so these power law corrections will typically be further suppressed in the chiral power counting.  We would like to further comment regarding the impact on the extraction of phase shifts by the use of the 
L\"{u}scher method~\cite{hamber_et_al,luscher_1,luscher_2}, since the main application we have in mind for our construction is the study of nuclear forces. In the L\"{u}scher method, the energy levels of a two-particle system are related to the (infinite volume) phase shift. In the case of boxes sizes $L$ much larger than the scattering length $a$ between the particles this relation is
\beq\label{eq:luscher}
 M\Delta E = \frac{4\pi a}{L^3} \left[1 + c_1 \frac{a}{L} + c_2 (\frac{a}{L})^2+ \cdots  \right],
 \eeq 
where $c_1$ and $c_2$ known numerical constants, $\Delta E=E-2M$ the shift in energy between the non-interacting and the interacting two-particle state and $M$ the mass of each particle.  With spatially periodic boundary conditions, this formula is exact up to corrections suppressed by $\sim e^{-m_\pi L}$~\cite{Bedaque:2006yi}. The ellipsis in equation (\ref{eq:luscher}) includes corrections due to the mixing of partial waves, starting at order $1/L^6$.  In the case of the orbifold boundary conditions, the first two terms in equation~(\ref{eq:luscher}) are unchanged since the power law corrections to the energy levels described by it are due to the on-shell propagation of nucleons whose boundary conditions are the same as in the periodic case.  The large ($\sim 1/L^3$) finite volume corrections to the mass of each particle cancel against the same correction to the energy levels of the two-particle state and does not affect the extraction of the scattering length.  In other words, the finite volume corrections to $E$ are cancelled by those of $2M$~\cite{Sato:2007ms}.  There will be, however, corrections to the nucleon-nucleon interaction (and consequently to the value of $a$) that are suppressed by $1/L^3$ only. However, combining it with the $1/L^3$ factor present in equation (\ref{eq:luscher}), the finite volume correction to the energy level will appear only at order $1/L^6$. This is parametrically smaller than several other corrections coming from contamination from higher partial waves, etc.. and are not a cause for concern, at least as long the condition $a\ll L$ is valid.  The breaking of hypercubic invariance (due to the special role of the $z$ direction) is more cause for concern as it allows for a mixing of higher partial waves beginning at order $1/L^5$.

\section{Conclusions}
We spelled out in detail an algorithm to implement orbifold boundary conditions in lattice QCD that eliminates the pion zero mode and, consequently, improves the signal-to-noise ratio in multi-baryon correlators. Our construction involves the use of a sea sector with periodic boundary conditions (and the use of standard, ``off-the-shelf" gauge configurations) and a valence sector with orbifold boundary conditions. In the most favored incarnation of our construction (the ``three-dimensional orbifold") the energy of the lowest pion mode is given by $\sqrt{9(\pi/L)^2+m_\pi^2}$ what is enough to completely eliminate the exponential growth of error in two-nucleon correlators for pion masses larger than about $m_\pi\approx 300$ MeV and lattice sizes $L\approx 4/m_\pi$ (and drastically reduce it for lighter pions). We pointed out that the absence of the pion zero mode enhances finite volume corrections, such that they are of order $\sim 1/L^3$, as opposed to $~e^{-m_\pi L}$ as in the periodic case. These enhanced finite volume effects, however, do not preclude the use of the L\"{u}scher formula to extract baryon-baryon phase shifts. Indeed, the corrections to the scattering length extracted through the L\"uscher formula are higher order in powers of $1/L$ and are furthermore suppressed in the chiral expansion.

  
\end{document}